\documentclass[aps,twocolumn,showpacs,superscriptaddress,floatfix,nofootinbib]{revtex4-1}

\usepackage{textcomp}
\usepackage{graphicx,graphics}
\usepackage{amsmath}
\usepackage{amssymb}
\usepackage{dsfont}

\usepackage{color}
\usepackage[normalem]{ulem}

\begin{document}
\title{Kinetic samplers for neural quantum states}

\author{Andrey A. Bagrov}
\email{andrey.bagrov@physics.uu.se}
\affiliation{Department of  Physics  and  Astronomy,  Uppsala  University, Box 516,  SE-75120  Uppsala,  Sweden}
\affiliation{Theoretical Physics and Applied Mathematics Department, Ural Federal University, 620002 Yekaterinburg, Russia}

\author{Askar A. Iliasov}
\email{a.iliasov@science.ru.nl}
\affiliation{Institute for Molecules and Materials, Radboud University, Heyendaalseweg 135, 6525AJ Nijmegen, \mbox{The Netherlands}}
\affiliation{Space Research Institute of the Russian Academy of Science, Moscow, 117997, Russia}

\author{Tom Westerhout}
\email{tom.westerhout@ru.nl}
\affiliation{Institute for Molecules and Materials, Radboud University, Heyendaalseweg 135, 6525AJ Nijmegen, \mbox{The Netherlands}}

\begin{abstract}
Neural quantum states (NQS) are a novel class of variational many-body wave functions that are very flexible in approximating diverse quantum states. Optimization of an NQS ansatz requires sampling from the corresponding probability distribution defined by squared wave function amplitude. For this purpose we propose to use kinetic sampling protocols and demonstrate that in many important cases such methods lead to much smaller autocorrelation times than Metropolis-Hastings sampling algorithm while still allowing to easily implement lattice symmetries (unlike autoregressive models). We also use Uniform Manifold Approximation and Projection algorithm to construct two-dimensional isometric embedding of Markov chains and show that kinetic sampling helps attain a more homogeneous and ergodic coverage of the Hilbert space basis.
\end{abstract}

\maketitle








The concept of neural quantum states (NQS) emerged several years ago, when it was suggested that variational wave functions possessing structure of simple neural networks -- restricted Boltzmann machines -- can be efficiently optimized to approximate ground states of some many-body quantum systems \cite{Carleo:2016}. The idea of using an ansatz of that type turned out to be very appealing because of neural networks' flexibility in representing data: instead of constructing a very specific trial function that accounts for physical properties of the concrete model of interest \cite{mvmc}, one could hope to get away with a universal neural approximator \cite{approximator} that can automatically adjust itself over the course of learning and approach ground state of {\it any} local Hamiltonian. However, soon it became clear that fermionic systems \cite{Clark} (away from the neutrality point) and frustrated quantum magnets are challenging for the NQS approach \cite{Neupert-frust, Westerhout:2020}, just as they are for the more traditional and established methods \cite{frust_book}. This posed a natural quest for improving upon this approach and bringing it closer to the point when it can be successfully applied to studying such models. Since then, the method of NQS has evolved into a solid framework embracing a number of optimization schemes and variational ans\"atze (going far beyond the originally proposed shallow Boltzmann machines) \cite{Hybrid1, Hybrid2, Hybrid3}, and considerable progress has been made in understanding both strong points and shortcomings of neural network wave functions \cite{Castelnovo}. On the positive side, it was realized that even the simplest NQS could host volume law entanglement \cite{Deng-Li, Levine-Sharir} and, in fact, have great capacity to neatly express a vast variety of many-body states, including ground states of frustrated spin Hamiltonians \cite{Westerhout:2020}. For instance, choosing a suitable NQS architecture combining the flexibility of a neural network with some prior knowledge about the model allowed to attain high accuracy in solving the $J_1-J_2$ Heisenberg antiferromagnet on square lattice and reveal a novel spin liquid phase of its ground state \cite{Nomura_liquid, Nomura_symmetry}. Some of the reasons why NQS could not be blindly applied to highly frustrated systems have been identified as well. The progress made in the field encourages further improvement of the method to make it suitable for studying many-body systems that are currently beyond its scope of applicability.

An important aspect of all the NQS optimization algorithms is Monte Carlo sampling. Since neural network architectures are not amenable to full contraction, computing loss functions (energies, fidelities) requires sampling from the probability distribution defined on the Hilbert space basis by the wave function amplitudes. At this point, exceptionally high expressibility of NQS, while being a clear advantage of the method, turns out to hold a hidden danger. During the learning procedure, NQS undergoes a sequence of weight updates, and the corresponding probability distribution evolves in a highly non-trivial way. It could easily happen that the distribution acquires a form which is problematic to sample from by means of Monte Carlo techniques. For example, if the distribution constitutes a number of well-separated narrow peaks on the set of basis vectors, inaccurate sampling could lead to ergodicity problems, incorrect estimates of the distribution, and, as a result, the NQS following a wrong direction on the optimization landscape.

Perhaps, the most promising way to overcome the non-ergodicity issue is to employ a certain class of neural architectures called generative models \cite{Salakhutdinov}, as was recently suggested. The most well-known example of generative models are autoregressive models \cite{Gregor}. These models are constructed to represent probability distributions as products of conditional probabilities. In the context of finite-dimensional lattice quantum models, the conditional probabilities have the meaning of probabilities for a subset of degrees of freedom, e.g. spins, to be in a certain classical state given the state of the complementing degrees of freedom fixed. Such representation allows to sample from the distribution {\it exactly} without resorting to Markov Chain Monte Carlo (MCMC) techniques \cite{Carleo:autoregressive}. The downside is that implementation of symmetries becomes problematic. So far, only the basic constraints such as the fixed total magnetization \cite{Carrasquilla_autoreg, Carrasquilla_U1} and translation invariance \cite{Roth} have been formulated within the framework of generative models. 

Since using all the accessible symmetries, such as the lattice symmetries, provides an essential advantage in studying many-body quantum systems \cite{Nomura_symmetry}, it is natural to ask whether there is an alternative way to bypass the problem of correlated samples generated with MCMC. In this paper, we propose an approach to sampling from NQS probability distributions based on the concept of continuous-time kinetic Monte Carlo. The idea behind it is to substitute the discrete chain of ``proposition-acceptance'' steps with a rejection-free process evolving in continuous time \cite{Zanella:2019}. Although well-appreciated in many other domains of computational physics \cite{Voter}, it has not been used within the domain of machine learning for quantum simulations, and here we make a step in this direction. In particular, we focus on the minimal continuous-in-time sampling algorithm which we shall call Zanella process following \cite{PowerGold:2019}. We consider several classes of many-body quantum states such as exact ground states of frustrated systems of up to 36 spins ($J_1-J_2$ Heisenberg antiferromagnet on square lattice, and the nearest-neighbor Heisenberg antiferromagnet on Kagome lattice), and neural quantum representations obtained during ground state optimization for the same models. For each state, we assess the quality of sampling protocols. The two gauges we use are autocorrelation time and the visualized coverage of configuration space constructed with Uniform Manifold Approximation and Projection (UMAP) dimension reduction algorithm \cite{UMAP}.

The paper is organized as follows. In Sec. \ref{sec:symmetries} we outline the implementation of lattice symmetries. In Sec. \ref{sec:KMC}, we provide a pedagogical introduction to Zanella process closely following Ref. \cite{PowerGold:2019}. Sec. \ref{sec:Results} contains the main results regarding the use of sampling protocols for different quantum states. We conclude with Sec. \ref{sec:Conc}.
\vspace{-0.17cm}
\section{Implementation of lattice symmetries} \label{sec:symmetries}
In the context of NQS, the conventional way to take into account symmetries of the lattice is to impose the corresponding constraints on the architecture of neural networks \cite{Neupert:2018}. While it is rather straightforward to implement translation invariance of a chain or a square lattice in this way, for more general (especially non-Abelian) point crystal symmetries, this approach becomes problematic. Instead, one can resort to operating within symmetry-adapted Hilbert space basis, which is often used in exact diagonalization \cite{Wietek:2018} (see also \cite{Symmetries, GroupSymmetry}). In our study, we adopt this approach and for completeness outline it here. We work under the assumption that the symmetry group has at least one one-dimensional irreducible representation, and the state of interest which one is sampling from, can be expressed as a combination of basis vectors from one of these representations.

Let $\mathcal{H}$ be the Hamiltonian and $A$ be the finite symmetry group generated by the lattice \emph{symmetry operators} $\{T_k \}$, i.e. operators which commute with the Hamiltonian: $[\mathcal{H}, T_k] = 0$. In $\sigma^z$ product state basis, spin configurations are represented as binary sequences $|\sigma\rangle=|\sigma_1\sigma_2 \ldots\sigma_n\rangle$, $\sigma_i=0,1$. 
To define the symmetry-adapted basis, we first introduce equivalence classes of basis spin configurations under the group action as orbits $\mathtt{orbit}(|\sigma\rangle) = \left\{g|\sigma\rangle | g \in A\right\}$. Every orbit is then represented by the basis state $|\tilde{\sigma}\rangle \in \mathtt{orbit}(|\sigma\rangle)$ which has the minimal value when viewed as a binary representation of an integer number: $\mathtt{representative}(|\sigma\rangle) = |\tilde\sigma\rangle = \min_\text{int} \mathtt{orbit}(\sigma)$. For example, orbit of basis vector $|\sigma\rangle=|\downarrow\downarrow\uparrow\uparrow\rangle \simeq \{0011\}=2^2+2^3=12$ in a periodic 4-spin chain would be represented by $|\tilde{\sigma}\rangle=|\uparrow\uparrow\downarrow\downarrow\rangle \simeq \{1100\}=2^0+2^1=3$.

To build the one-dimensional representation, we note that, since $A$ is finite, for every symmetry generator $T_k$, there is a $n_k$ such that $T_k^{n_k} = \mathds{1}$, which is typically quite small (at most of the order the of system size).
Hence, eigenvalues of each symmetry generator are roots of $1$, and $T_k|0\rangle =\lambda_k|0\rangle$, $|\lambda_k|=1$, where $|0\rangle$ is the ground state of $\mathcal{H}$. Thus for any $g \in A$, one can write $g|0\rangle = \lambda_g |0\rangle$, and $\lambda_{gh} = \lambda_{g} \lambda_{h}$ for $g, h \in A$, which determines the one-dimensional irreducible representation of the symmetry group.
Importantly, even if $A$ itself is a non-Abelian group, this construction is valid as long as its representation is Abelian. Although 
generally $[T_i,T_j]\neq 0$, on the ground state (as well as any other state $|\sigma\rangle$ belonging to this representation): $[T_i,T_j]|\sigma\rangle=0$.

For each $|\sigma\rangle$, there exists a $g\in A$ such that $\langle \sigma |0\rangle = \langle \tilde\sigma |g^\dagger |0\rangle = \lambda_g^* \langle\tilde\sigma |0\rangle$. Thus 
\begin{equation}
    \langle \sigma |0\rangle \cdot |\sigma\rangle = \lambda_g^* \langle\tilde\sigma |0\rangle \cdot g|\tilde\sigma\rangle =  \langle\tilde\sigma |0\rangle \cdot \lambda_g^* g |\tilde\sigma\rangle. 
\end{equation}
This means that the standard basis expansion of $|0\rangle$ can be rewritten as
\begin{gather}\label{eq:basis_expansion}
    |0\rangle = \sum_{\sigma} \langle \sigma| 0\rangle \cdot |\sigma\rangle
              = \sum_{\tilde\sigma} \sum_{g\in A} \frac{N_{\tilde\sigma}}{|  A|}\langle \tilde\sigma| 0\rangle \cdot \lambda_g^* g |\tilde\sigma\rangle = \\ \nonumber \frac{1}{|A|}\sum_{\tilde\sigma} N_{\tilde\sigma} \langle \tilde\sigma| 0\rangle \cdot \sum_{g\in A} \lambda_g^* g |\tilde\sigma\rangle,
\end{gather}
where $N_{\tilde\sigma} \in \mathbb{N}$ is the number of original basis elements in the orbit of $|\tilde\sigma\rangle$; $|A|$ denotes the number of elements in the symmetry group $A$; the sum over $\sigma$ runs over all basis vectors of the Hilbert space, and the sum over $\tilde\sigma$ runs over representatives of all orbits. Using \eqref{eq:basis_expansion}, we define a new basis:
\begin{equation}\label{eq:symm_basis}
    |\mathcal{S}_{\tilde\sigma}\rangle = \frac{1}{\sqrt{N_{\tilde\sigma}}} \sum_{g\in A} \lambda_g^* \cdot g |\tilde\sigma\rangle \,,
\end{equation}
where $1/\sqrt{N_{\tilde\sigma}}$ coefficient is introduced to ensure proper normalization.

We can redefine the Hamiltonian $\mathcal{H}$ in the new basis. Suppose that originally we had $\mathcal{H}|\sigma\rangle = \sum_i c_i |\sigma_i\rangle$. Then in the symmetry-adapted basis we get:
\begin{equation}\label{eq:hamiltonian_action}
    \begin{aligned}
        \mathcal{H}|\mathcal{S}_{\tilde\sigma}\rangle=
            \sum_i c_i \frac{\sqrt{N_{\tilde{\sigma}_i}}}{\sqrt{N_{\tilde\sigma}}} \lambda_{h_i} \cdot |\mathcal{S}_{\tilde{\sigma}_i}\rangle.
    \end{aligned}
\end{equation}

One should keep in mind that, when constructing the symmetry-adapted basis, we used the ground state $|0\rangle$ for illustrative purposes only, to make sure that the representation we are dealing with includes $|0\rangle$. In a real world scenario, one does not know the ground state beforehand, and in fact it is not required to find characters $\lambda$ of the representation. If the group has several one-dimensional irreducible representations, and it is unknown which one the ground state belongs to, the optimization problem should be solved separately in each of the corresponding symmetry-adapted bases.

\section{Kinetic Monte Carlo} \label{sec:KMC}

Metropolis-Hastings algorithm is the most popular choice for Markov chain sampling from a probability distribution $\pi(x)$ defined on a discrete set of elements ${\cal X}$, such as the Hilbert space basis of a finite-dimensional quantum system.
At every iteration, state of the chain is given by some $x\in {\cal X}$. An element $y$ from the vicinity $\partial x$ of $x$ is then suggested, and the sampling process either transitions to $y$ or remains at $x$. The transition happens with probability $p=\min(1,\frac{\pi(y)}{\pi(x)})$.
Which elements should be considered as belonging to $\partial x$ depends on the problem, but for closed quantum spin systems with fixed magnetization, where every element is a product state $|\uparrow \downarrow \uparrow \dots \downarrow \downarrow \rangle$, $\partial x$ is often chosen to include elements that differ from $x$ by a binary spin flip that preserves total magnetization.

If, for some $x$ it turns out that  $\pi(x) \gg \pi(y)$ for the majority of $y \in \partial x$, acceptance rate becomes very low, and the sampling process gets stuck at $x$ for many iterations. This negatively affects the quality of sampled sequence making it too correlated and not accurately representing the desired $\pi(x)$ distribution. If $\pi(x)$ has a number of far-separated peaks of this kind, or if elements of $\cal X$ tend to form clusters such that the Markov chain cannot leave them once entered, the sampling could lead to severely wrong results.

Zanella process \cite{Zanella:2019, PowerGold:2019} is a natural way to bypass this problem by using a rejection-free scheme instead of the ``acceptance-rejection'' protocol. As before, at every step the sampling process is located at some $x_i\in {\cal X}$. Now however, even if this point is a local maximum of the probability distribution and acceptance rate in Metropolis-Hastings algorithm would be very low, the process still jumps to an $x_{i+1}$ from $\partial x_i$. To preserve the information about probability distribution we introduce a waiting time. In other words, before jumping, the process sits at $x_i$ for some time $\tau_i \in \mathbb{R}$ which depends on the probability ratio $\pi(x_{i+1})/\pi(x_i)$. Hence, instead of getting stuck at $x_i$ for many steps, $\tau_i$ is set to a high value and the process moves on. To be more precise, the algorithm can be outlined as follows:
\begin{itemize}
    \item At step $i$, compute normalization for the probability of jumping away from $x_i$ (i.e. a ``decay rate'')
    \begin{equation}
        \lambda_{i} = \sum\limits_{y\in\partial x_i}g\left(\frac{\pi(y)}{\pi(x_i)}\right), \nonumber
    \end{equation}
    where $g$ is a function obeying $g(l) = l\cdot g(1/l)$, usually called a \emph{balancing function} \cite{PowerGold:2019}.
    \item Estimate waiting time $\tau_i$ by sampling it from the exponential distribution
    \begin{equation*}
        \tau_i \sim \mathrm{Exp}(\lambda_i) \;,
    \end{equation*}
    where the probability density function (PDF) of $\mathrm{Exp}(\lambda)$ is $f(x;\lambda) = \lambda e^{-\lambda x}$.
    \item Increase the overall running time
    \begin{equation}
        t_{i+1} = t_i+\tau_i \,. \nonumber
    \end{equation}
    \item Choose a state $y\in \partial x_i$ with probability
    \begin{equation}
        p(y) = \frac{1}{\lambda_i}\cdot g\left(\frac{\pi(y)}{\pi(x_i)}\right)\,. \nonumber
    \end{equation}
    \item Jump to $x_{i+1} = y$ and repeat the scheme.
\end{itemize}
In this formulation, one can think of the sequence $\{x_i\}$ of samples as if it were a piece-wise constant function of time $x(t)$. Expectation value of a function defined on $\cal X$ can then be computed as follows:
\begin{gather}
    E_\pi \left[f(x)\right] = \lim\limits_{T\rightarrow \infty}\frac{1}{T}\int\limits_{0}^{T} f(x(t))dt =\\ \lim\limits_{N\rightarrow \infty} \sum\limits_{k=0}^{N-1} \frac{t_{k+1} - t_k}{t_N}f(x_k)
                       \simeq \lim\limits_{N\rightarrow \infty}\frac{1}{N}\sum\limits_{k=0}^{N-1} f(k\Delta t), \nonumber
\end{gather}
where $\Delta t = t_N / N$. Balancing function $g$ plays a role similar to the importance sampling in MCMC, but in this paper we consider the simplest case of $g=1$.

As we will see, already this simple algorithm allows to drastically improve ergodicity of sampling from ``problematic'' distributions. 

\section{Kinetic sampling vs. Metropolis-Hasting} \label{sec:Results}
\subsection{Assessment criteria}
The standard way to estimate the quality of Monte Carlo sampling is to compute autocorrelation time $\tau_{corr}$ for some relevant quantity $O$. Autocorrelation time is the characteristic decay time (number of steps in the Markov chain) of the corresponding two-point correlation function \cite{Sokal}:
\begin{gather*}
    C_{O}(t) = \langle O(t) O(0)\rangle - \langle O(t) \rangle^2, 
    \\
    C_O(t)/C_O(0)\simeq e^{-t/\tau_{corr}},\text{ for } t \gg 1
\end{gather*}
In the context of sampling from probability distributions given by many-body wave functions, two natural choices of $O$ are logarithmic probability $\log |\psi({\cal S})|^2$ and local energy estimator $E_{loc}({\cal S})$ defined by the following equation:
\begin{gather*}
    E = \langle\psi| H |\psi\rangle
      = \sum_{\cal S} \frac{\langle {\cal S} | H | \psi\rangle}{\langle {\cal S} | \psi\rangle} \cdot |\langle {\cal S} | \psi\rangle|^2
      \equiv \\ \sum_{\cal S} E_{loc}({\cal S}) \cdot |\langle {\cal S} | \psi\rangle|^2
      \approx \sum_{{\cal S}\sim |\psi|^2} E_{loc}({\cal S}) \;.
\end{gather*}
Logarithmic probability is chosen instead of $|\psi({\cal S})|^2$ to avoid numerical issues. In the following, we will use $C_\psi(t)$ to denote autocorrelation function computed for $\log|\psi({\cal S}(t))|^2$, and $C_E(t)$ --- for $E_{loc}({\cal S}(t))$.

Although autocorrelation time is a well-established criterion, it is not the only way to judge the quality of drawn samples. In \cite{Carleo:autoregressive}, authors used PCA \cite{PCA} algorithm to reduce the dimension
of basis vectors. By visualizing the resulting 2D vectors, they were able to compare Metropolis-Hastings algorithm to exact sampling procedure of autoregressive neural network architectures and demonstrate that the latter leads to much more homogeneous coverage of the Hilbert space basis. For systems considered in this work, PCA does not seem to reveal much additional information about the quality of samplers. Thus we suggest to use a more involved but arguably also superior dimension reduction algorithm called UMAP (Uniform Manifold Approximation and Projection) \cite{UMAP}. We refer the reader to the original paper for a detailed motivation and description of the algorithm (especially Sec. 3), but would still like to briefly outline the procedure here.

UMAP algorithm operates on a discrete dataset $\left\{x_i\right\}$ equipped with some metric $d(x_i, x_j)$ which measures dissimilarity between data points. In our case, the dataset is a subset of Hilbert space basis sampled using either Metropolis-Hastings or Zanella algorithm (excluding all duplicates). The metric is simply the Hamming distance between the corresponding spin sequences $|\uparrow \downarrow \dots \uparrow\uparrow \rangle$. The first stage of the algorithm is to equip the dataset with a structure of a weighted undirected graph. One fixes the number of nearest neighbours $k$ every vertex (basis vector) should be connected with. For our purpose, it is natural to choose $k\lesssim N$, where $N$ is the number of spins in the system. Concretely, for 36-spin systems we take $k=10$. In the resulting graph, each edge is assigned some weight $w_{ij}$ which is a function of $d$ and $k$. Once the graph is constructed, UMAP algorithm projects it on a low-dimensional space (usually the 2D plane) in a way that approximately preserves the distances between the vertices, such that the resulting visualization maximally accurately represents the actual metric structure of the dataset.
Embedding into a 2D space allows to directly compare quality of different samplers by contrasting sampled sequences visually for different types of many-body quantum states. 

To represent Monte Carlo samples we adopt the following protocol:
\begin{itemize}
    \item Using each of the two samplers (Metropolis-Hastings and Zanella), generate a sequence of $10^5$ vectors.
    \item Merge these two sequences and discard all repetitions to obtain a set of unique basis vectors visited by either of the samplers.
    \item Equip this set with Hamming distance and build its UMAP embedding into the two-dimensional plane.
    \item Visualize the sequences within this embedding.
\end{itemize}

\subsection{Benchmarks}
\begin{figure*}[t!]
    \centering
    \includegraphics[width=0.9\textwidth]{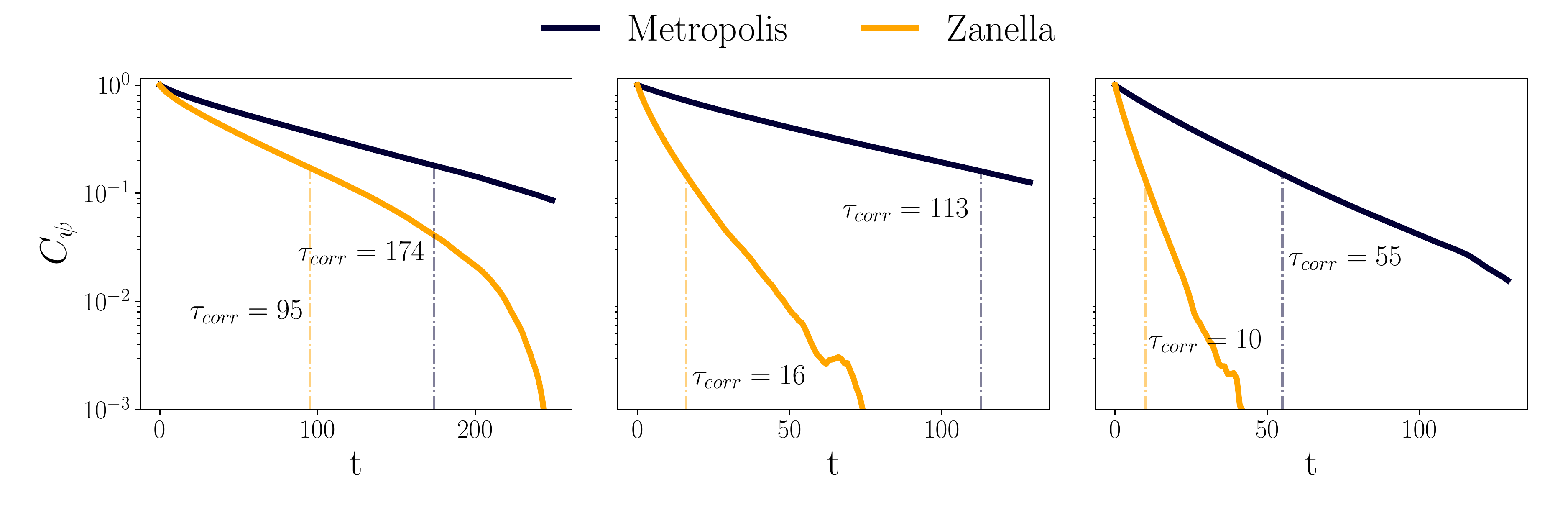}
    \vspace{-0.5cm}
    \caption{\label{fig:GS_autocorr}Autocorrelation function $C_\psi$ of Metropolis-Hastings and Zanella processes computed for the cases of sampling from probability distributions corresponding to ground states of Heisenberg antiferromagnet on square lattice at $J_2=0$ (left) and $J_2=0.55$ (middle) and Kagome lattice (right). Autocorrelation function was computed by averaging $300$ chains of length $8000$.}
\end{figure*}

First, we compare the Monte Carlo algorithms by sampling from the exact ground states of three quantum spin models: the $J_1-J_2$ Heisenberg antiferromagnet on $6$-by-$6$ square lattice with periodic boundary conditions at $J_2=0.0$ (non-frustrated) and $J_2=0.55$ (maximally frustrated), and Heisenberg antiferromagnet on $36$-site Kagome cluster with periodic boundary conditions. For exact ground states, energy autocorrelation functions are not well-defined because $E_{loc}(\mathcal{S})$ are identical for all basis spin configurations $\mathcal{S}$. We thus only compute the probability autocorrelation functions $C_\psi$ which are shown in Fig.~\ref{fig:GS_autocorr}. For all three systems, Zanella process shows superior performance. In the ground states, it allows to reduce autocorrelation time by a factor of 2--7, and, as will be shown below, for more generic states encountered during NQS optimization the gain could be up to a factor of 10--50. We expect this effect to increase even more for larger systems.
\begin{figure*}[t!]
    \includegraphics[width=0.9\textwidth]{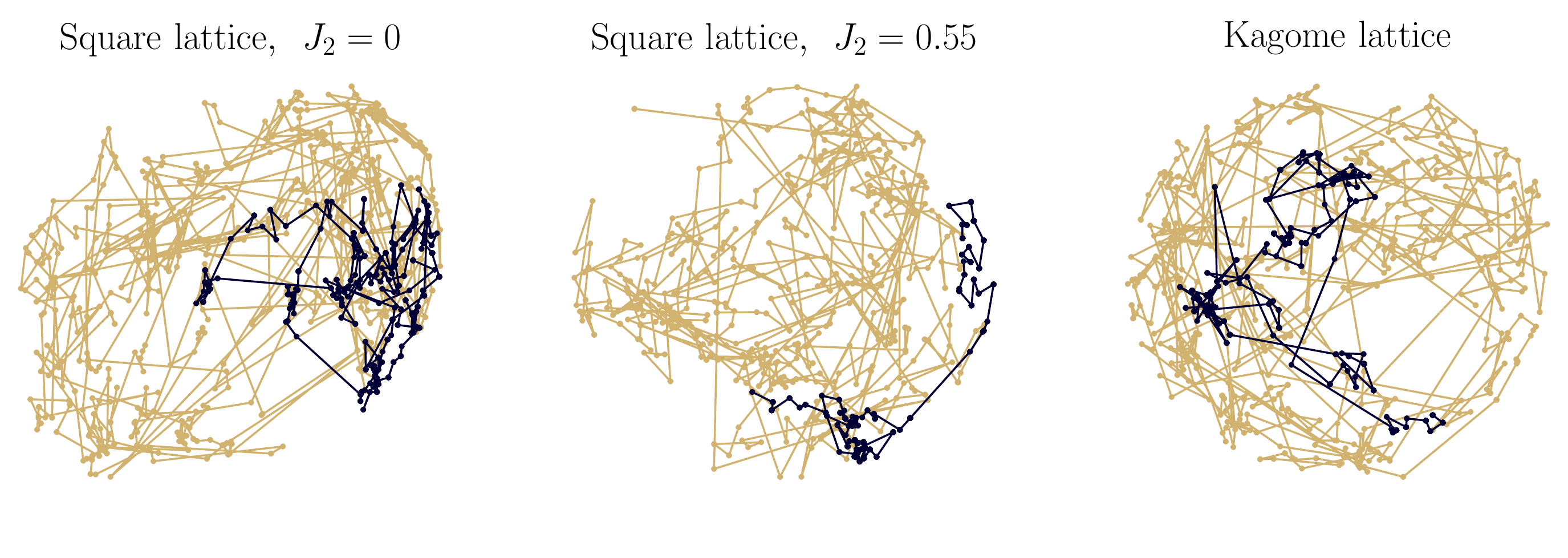}
    \vspace{-0.5cm}
    \caption{\label{fig:UMAP}UMAP visualization of sequences of basis elements produced by Metropolis-Hastings (dark blue) and Zanella (tan) samplers. The sampled probability distributions correspond to ground states of Heisenberg antiferromagnets on square lattice at $J_2=0$ and $J_2=0.55$ and Kagome lattice. Every point represents a vector from the non-symmetrized basis. For both samplers 800 elements are shown.}
\end{figure*}

Secondly, we analyze Zanella and Metropolis-Hastings algorithms using UMAP dimension reduction. Defining a metric on the symmetry-adapted basis is a non-trivial task and is left for future studies. Instead, we do the sampling in non-symmetrized basis. Dimension of the Hilbert space is thus $9075135300 = \mathcal{O}(10^{10})$ for all three systems. Hamming distance acquires a very concrete physical interpretation --- it counts the minimal number of steps an algorithm needs to move from one basis state to another. In Fig.~\ref{fig:UMAP} we show relatively short (800 steps) parts of Markov chains. They are taken from the middle of the chain to ensure that thermalization effects do not disrupt the picture. Visual distance in the figure approximately reflects the Hamming distance between points (of course, some aberrations caused by embedding into a lower-dimensional space are unavoidable). One can see that for all three considered wave functions the kinetic sampler explores the Hilbert space in a much more ergodic and swift manner than Metropolis-Hastings algorithm. Note also that for square lattice Metropolis-Hastings algorithm samples more unique states and covers a bigger part of the basis for $J_2=0$ than it does for $J_2=0.55$, even though its autocorrelation time is much larger in the former case. Without UMAP algorithm one would have ended up under the impression that the frustrated case was easier to sample.
\begin{figure*}[t]
    \includegraphics[width=0.9\textwidth]{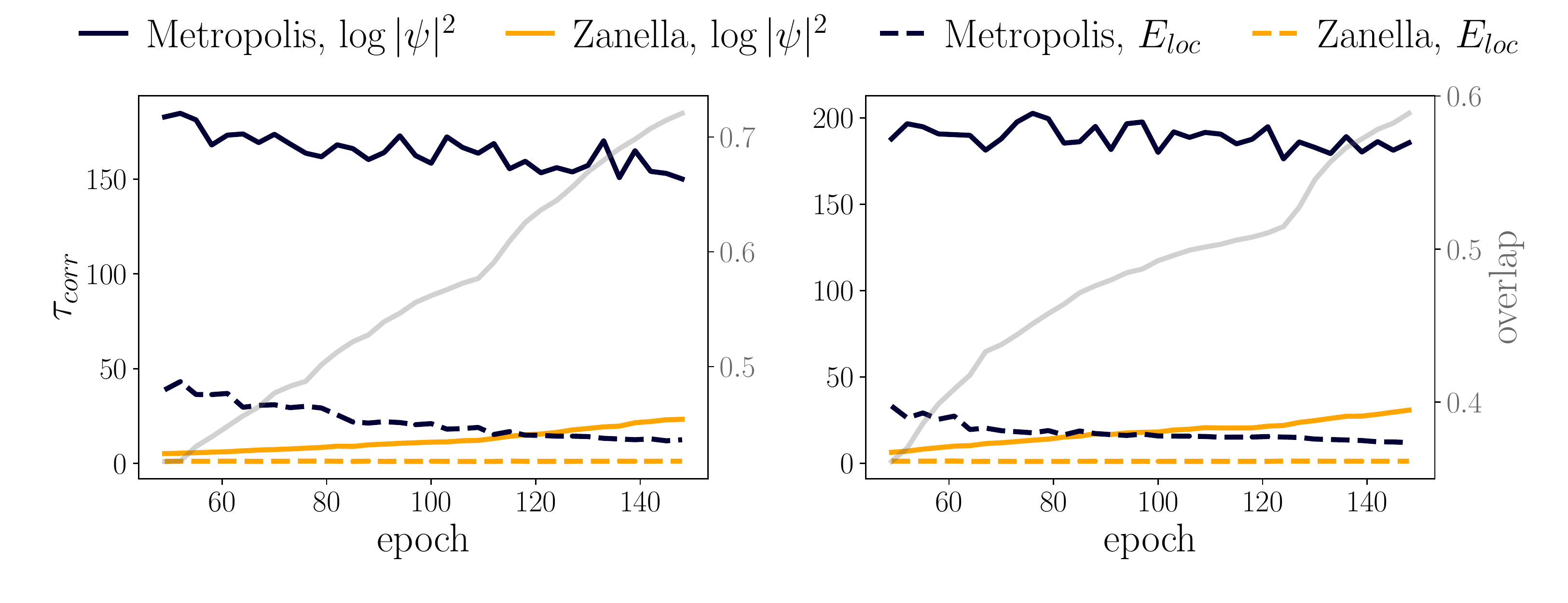}
    \vspace{-0.5cm}
    \caption{\label{fig:6x6_autocorr}Evolution of autocorrelation time during the NQS training procedure for Heisenberg antiferromagnet on 6-by-6 square lattice with $J_2=0$ (left) and $J_2=0.55$ (right). Autocorrelation times were estimated from $200$ chains of length $7000$. Upon approaching the ground state, the advantage of Zanella process becomes less significant, but over the course of learning it outperforms Metropolis-Hasting sampling by at least an order of magnitude.}   
\end{figure*}
\begin{figure*}[t]
    \includegraphics[width=0.9\textwidth]{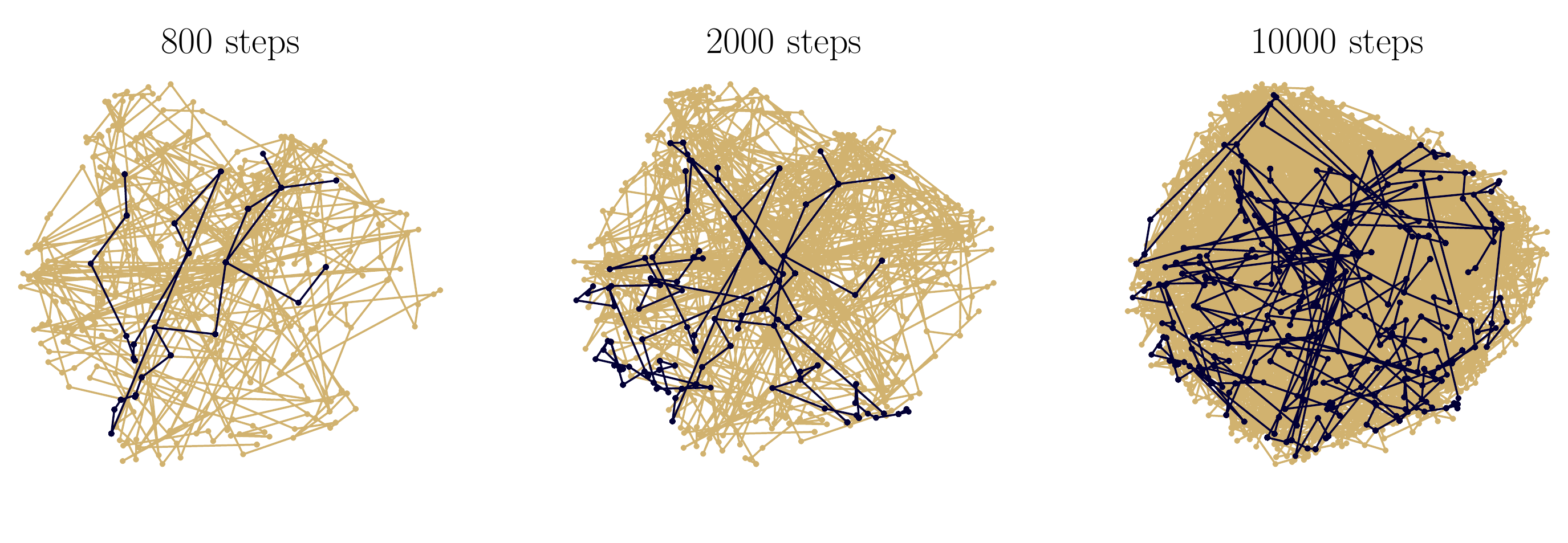}
    \vspace{-0.5cm}
    \caption{\label{fig:umap_nqs}UMAP visualization of sequences of basis elements produced by Metropolis-Hastings (dark blue) and Zanella (tan) samplers. The sampled probability distribution corresponds to a typical undertrained NQS (Heisenberg antiferromagnet on 6-by-6 square lattice with $J_2=0.55$, epoch \textnumero $\,86$). Every point represents a vector from the non-symmetrized basis. As before, the parts of Markov chains are taken from the middle of the chain to ensure proper thermalization.}   
\end{figure*}

Since we are mainly motivated by improving sampling in the context of NQS, it is instructive to compare quality of the methods in a realistic learning scenario. To perform this test, we run Stochastic Reconfiguration optimization \cite{Carleo:2016, Sorella} of simple neural-networks in the symmetry-adapted basis aiming at finding good approximations to ground states of the $J_1-J_2$ model on a $6$-by-$6$ square lattice with periodic boundary conditions at $J_2=0$ and $J_2=0.55$. Kagome lattice is still beyond the scope of applicability of the NQS method and we do not consider it here. We represent absolute values and signs of the wave function coefficients with two independent 1-hidden-layer dense networks with $4\cdot 36 = 144$ hidden neurons. To make the comparison unbiased, to optimize the NQS, we use neither of the Monte Carlo samplers, but rather compute energies of the variational states and the weight updates at every epoch by means of exact sampling. We view squared amplitudes of the wave as a discrete probability distribution and sample from it directly using standard textbook algorithms \cite{Walker:1974}. After the optimization, we go through obtained neural quantum states at every epoch of training and sample from them using Metropolis-Hastings and Zanella algorithms. Corresponding autocorrelation times are shown in Fig.~\ref{fig:6x6_autocorr}. One can see that autocorrelation times computed from both energy and probability correlation functions $C_E$ and $C_\psi$ are significantly smaller for the Zanella process. Upon approaching the ground state, ``probabilistic'' autocorrelation time of the Zanella process tends to increase, but using kinetic sampling remains highly advantageous at all stages of optimization. In Fig.\ref{fig:umap_nqs}, we show UMAP visualization of Metropolis-Hastings and Zanella Markov chains in the case of sampling from a typical NQS encountered during the learning process. The advantage of kinetic sampling over the Metropolis-Hastings algorithm is evident. \section{Conclusions} \label{sec:Conc}
In most of the NQS optimization algorithms, unless one is using generative models, Monte Carlo sampling is required to compute observables and gradients which makes it a crucial part of the learning scheme. In this paper, we have analyzed how the quality of sampling from probability distributions defined by many-body wave functions can be improved by using a kinetic Monte Carlo algorithm, --- continuous-in-time Zanella process, --- instead of the conventional Metropolis-Hastings algorithm. Being extremely easy to implement, Zanella process gives a substantial improvement in autocorrelation times. To further assess the quality of sampling, we proposed to employ UMAP embedding algorithm which constructs visualizations of high-dimensional datasets approximately preserving distances between elements. It thus serves as a much better source of geometric intuition about the dataset structure than, for example, principal component analysis. As follows from UMAP analysis, on top of having smaller autocorrelation times, Zanella process gives a more uniform coverage of the Hilbert space basis.

Possibly, the main research domain where the advantage provided by kinetic sampling could be of high importance is NQS application to real-time dynamics of non-equilibrium quantum many-body systems. In settings of that kind, not only does the resulting quality of approximation matter, but every single step of the simulation should conform with energy-preserving Hamiltonian evolution. Slight non-ergodicity of the sampler could introduce deviations from the proper evolution trajectory that would eventually lead to accumulation of large errors. In this context, employing a sampling algorithm that generates high-quality uncorrelated sequences could be as important as using neural network architectures with good expressibility and generalization properties. 

Although even the simplest Zanella algorithm appears to be superior to Metropolis-Hastings algorithm, further improvements are possible. Using rejection-free continuous-in-time process allows to avoid getting trapped at the same point for many iterations. However, another possible danger is localization of Markov chain within a small subset of the space $\cal X$. If the process enters a region of high probabilities, it could start wandering along short closed trajectories within this region such as $x\rightarrow y \rightarrow z \rightarrow x \rightarrow \dots$, which would negatively affect ergodicity. For probability distributions of this kind, an algorithm that forbids back-tracking might be desirable.
Recently, an extension of Zanella process has been suggested which approximately avoids back-tracking on short-to-medium time scales. This is done by promoting Zanella process to a non-Markovian metaheuristic which combines ideas of kinetic Monte Carlo and self-avoiding walks. The new algorithm was named Tabu sampler \cite{PowerGold:2019}. When applied to sampling from probability distributions on large graphs, Tabu sampler was shown to decrease autocorrelation times by one or two orders of magnitude compared to Zanella process. Implementing it for sampling from many-body wave functions is straightforward if lattice symmetries of the quantum system are not taken into account. However, the algorithm requires non-trivial modifications to be applied in the symmetry-adapted basis, which is a direction for future research.
\vspace{-0.17cm}
\section*{Acknowledgements}
\vspace{-0.17cm}
Authors thank Olle Eriksson, Mikhail Katsnelson and Danny Thonig for useful discussions. The work of T.W. was supported by European Research Council via Synergy Grant 854843 - FASTCORR. A.A.I. acknowledges financial support from Dutch Science Foundation NWO/FOM under Grant No. 16PR1024. A.A.B. acknowledges support from the Russian Science Foundation, Grant No. 18-12-00185. This work was partially supported by Knut and Alice Wallenberg Foundation through Grant No. 2018.0060. This work was carried out on the Dutch national \mbox{e-infrastructure} with the support of SURF Cooperative.
\vspace{-0.5cm}


\end{document}